# Nutritional composition and bioactive compounds of mini watermelon genotypes in Bangladesh

**Hasina Sultana[a], Sharmila Rani Mallick[a], Jahidul Hassan[a], Joydeb Gomasta[a], Md. Humayun Kabir[b], Md. Sakibul Alam Sakib[c], Mahmuda Hossen[c], Muhammad Mustakim Billah[b], Emrul Kayesh[a,*]**

[a] *Department of Horticulture, Bangabandhu Sheikh Mujibur Rahman Agricultural University, Gzipur, Bangladesh*
[b] *Department of Soil Science, Bangabandhu Sheikh Mujibur Rahman Agricultural University, Gzipur, Bangladesh*
[c] *College of Agriculture Sciences, International University of Business Agriculture and Technology Dhaka, Bangladesh*

\* Corresponding author.
*E-mail address:* ekayeshhrt@bsmrau.edu.bd (E. Kayesh).

ABSTRACT

Given the present rising trends in changing lifestyle and consumption patterns, watermelon production has shifted from big to small-sized fruits having desirable quality attributes. Hence, analyses of fruit quality traits of mini watermelon are crucial to develop improved cultivars with enhanced nutritional compositions, consumer-preferred traits and extended storage life. In this context, fruit morphological and nutritional attributes of five mini watermelon genotypes namely BARI watermelon 1 ($W_1$), BARI watermelon 2 ($W_2$), L-32468 ($W_3$), L-32236 ($W_4$) and L-32394 ($W_5$) were evaluated to appraise promising genotypes with better fruit quality. The evaluated genotypes expressed different levels of diversity for fruit physical qualitative traits including differences in shape, rind and flesh color and texture. The study also revealed significant variability among the genotypes regarding all observed fruit morphological and nutritional aspects as well as bioactive compounds. Among the studied genotypes, $W_1$ stood out with the highest TSS as well as rind vitamin C and total phenolic content accompanied by higher fruit weight and thick rind. On the other hand, $W_3$ genotype was featured with higher amount of β carotene, total phenolic and flavonoid content in its flesh along with rind enriched with β carotene and minerals. However, comparatively higher amount of sugar and total flavonoid content was recorded in the rind of $W_5$ genotype. Therefore, BARI watermelon 1 and L-32468 could be exploited for table purpose and using in breeding program to develop mini watermelon cultivars with more attractive fruits in terms of quality acceptance and nutritional value in Bangladesh. Furthermore, rind of BARI watermelon 1 and L-32394 could be considered as the potential cheap source of bioactive compounds to be used for dietary and industrial purpose which would decrease the solid waste in the environment.

*Keywords:* Mini watermelon, Nutritional composition, Rind, TSS, TPC, TFC.

## 1. Introduction

Watermelon (*Citrullus lanatus*) is one of the most popular fruits of Cucurbitaceae family widely distributed in the tropics and sub tropic regions (Gladvin et al., 2017). It is largely consumed as refreshing summer fruit, much appreciated because of its refreshing capability, attractive color, delicate taste and high water content to quench the summer thirst (Asfaw, 2022). In the year 2020, a production area of 3.05 million hectares was employed for the production of 101 million tons of watermelon throughout the world with Asia, as a continent, contributing to 81% of the total production (Morales et al., 2023; Assefa et al., 2020). In Bangladesh, its cultivation has become more popular and profitable agribusiness in recent years due to its relatively higher yield per unit area among different kinds of fruits grown in the country (Rabbany et al., 2013).

For the commercial market, watermelon with good quality is always preferred by consumers. The external quality attributes of watermelon such as shape, weight, and rind and flesh colors are important preference components for consumers to purchase the fruit (Kyriacou et al., 2018). Recently, in addition to the external quality factors, consumers also consider the internal quality features such as the fruit's sugar and nutritional contents (Musacchi and Serra, 2018). As a nutrient-dense, low energy food, watermelon provides vital nutrients and contributes to overall fruit intake (Fulgoni and Fulgoni, 2022). Its flesh contains a large amount of water, which is approximately 93% of the total weight of the flesh (Liu et al., 2018) and also contains micronutrients such as vitamins, minerals, amino acids (citrulline and arginine), lycopene and bioactive compounds (Rico et al., 2020; Manivannan et al., 2020). Generally, watermelon rind is treated as agricultural waste and discarded after consuming the flesh causing environmental issues and biomass loss (Xiaofen and Ramirez, 2022). Though the rind is not as juicy as the flesh but it is edible and has many health benefits due to the presence of important amino acid citrulline, fiber, minerals and phenolic compounds (Mohan et al., 2016; Ashoka et al., 2022). Moreover, utilization of the rind as an ingredient has been studied in products including pickle, candy, cheese, etc. (Mohamed et al., 2013). Hence, it would be favorable to take advantage of the nutritional potential of rind and create commercial value, rather than limiting it to agricultural waste.

The economic and nutritional values of watermelon recognized in recent years have created the opportunity to develop and commercialize new varieties combining high fruit yield and quality (Yang et al., 2016). Nowadays, there have been changes in the population patterns that led to increasingly smaller families and, as a consequence, a preference for smaller fruits, such as the mini watermelon which weigh 2-4 kg. In addition, consumers having low incomes prefer small-to-medium-sized fruits rather than large fruits because of their high prices. As well as being easy to handle, they also occupy less space in refrigerators (Sari et al., 2016). Watermelon production is, therefore, ultimately shifted from big fruits to small-sized fruits having desirable quality attributes (Tegen et al., 2021). In Bangladesh, mini watermelons are also gaining popularity and to the best our knowledge, scientific information is scarce about their fruit quality and nutritional status. Hence, the present study was designed to assess the fruit quality of the mini watermelon genotypes in terms of their physical as well as nutritional properties for selecting promising genotypes to be used for future watermelon breeding programs in Bangladesh.

## 2. Materials and Methods

### 2.1. Experimental site

The experiment was conducted at the Department of Horticulture, Bangabandhu Sheikh Mujibur Rahman Agricultural University (BSMRAU), Gazipur-1706, Bangladesh during

February to August 2022. This experimental area belongs to the agro-ecological zone Madhupur Tract (AEZ 28) (24°09° N latitude and 90°26° E longitude; 8.4 m above sea level) having mean temperature varies from 28-32° C in summer season but winter season shows falling below 2° C and annual rainfall lies between 1000-1500 m. The soil was clay loam in texture and acidic in nature with a pH of around 5.8 (Khan et al., 2023).

*2.2. Plant material*

The seeds of two varieties and three lines of mini watermelon were collected from Bangladesh Agriculture Research Institute (BARI), Joydebpur, Gazipur-1701, Bangladesh and Lalteer Seed Company, respectively which were denoted by different accession number viz. $W_1$ (BARI watermelon 1), $W_2$ (BARI watermelon 2), $W_3$ (L-32468), $W_5$ (L-32236) and $W_6$ (L-32394).

*2.3. Plant growth*

Fresh, healthy and mature seeds were soaked in water for 3 hours and sown in February, 2022 in $4 \times 5$-inch polythene bag using three seeds each with garden soil and compost mix (1:1). At three to four true leaves stage, seedlings were transferred to the main research field following randomized complete block design (RCBD) replicated thrice with seven plants in each replication following a spacing of 1 x1 $m^2$. The plants at their subsequent growing stages were fertilized with proper doses of manures and fertilizers following Fertilizer Recommendation Guide of Bangladesh (FRG, 2018). Intercultural operations such as weeding, irrigation, mulching, trellising, pheromone trap setting, pesticide and fungicide spraying, etc. were done as per requirements.

*2.4. Fruit harvest and data Collection*

Five fully matured and ripe fruit per genotypes were randomly harvested in August 2022. Maturity was assessed by dried tendril, yellow ground spot, and hollow sound when fruits were tapped (Correa et al., 2020). Harvested fruits were immediately taken to the laboratory used for determinations of fruit physical and nutritional quality traits.

*2.5. Fruit physical attributes*

Fruit appearance was judged visually for fruit shape, rind pattern and flesh color while flesh texture was assessed by mouth feel. Fruit length was measured from the blossom end to the stem end, while fruit diameter was measured across the fruit between the blossom end and stem end. Rind thickness was estimated from the flesh to the outer rind of the fruit and taken at the midway point between the blossom and stem end on each side. The weight of each fruit was determined using a top pan electric balance.

*2.6. Nutritional attributes*

For nutritional quality estimation, fruit was prepared by washing with tap water, drying with paper towels and peeling with a kitchen knife.

*2.6.1. Total soluble solid*

Total soluble solid (TSS) content measurement of the flesh was performed with a hand refractometer (Model: Atago N1, Japan). A drop of flesh juice squeezed from 1g of sample was placed on the prism of the refractometer and the TSS content was recorded as degree Brix (°Brix) (Ranganna, 1995).

*2.6.2. Total sugar and reducing sugar*

Bertrand A, Bertrand B and Bertrand C standard solutions were used for the estimation of total and reducing sugar content in the flesh and rind of watermelon fruits according to the procedure described by Somogyi (1952).

*2.6.3. β carotene*

β carotene was analyzed as per Nagata et al. (1992). 1 g of flesh and rind samples were blended with mortar pestle, added to 10 mL of acetone: hexane (4:6) solution and mixed thoroughly. Then the sample was filtered with Whatman no. 42 (2.5µm particle retention) filter paper and optical density was measured by spectrophotometer (PD-303 UV Spectrophotometer; APEL Co.) at 663 nm, 645 nm, 505 nm, and 453 nm. Following formula was used for the calculation of β carotene (mg/100 g)

$$\beta \text{ carotene (mg/100 g)} = 0.216 \,(OD_{663}) + 0.452 \,(OD_{453}) - 1.22 \,(OD_{645}) - 0.304 \,(OD_{505})$$

Where, OD = optical density at particular wave length
0.216, 0.452, 1.22, 0.304 = absorption coefficient of the respective absorbance.

*2.6.4. Vitamin C*

Vitamin C as ascorbic acid was determined using titration method (Elgailani, 2017) with some modifications. 100 ml of sample extract was prepared by blending 20 g of sample with distilled water and centrifuging for 20 minutes at 4000 rpm with a centrifuge machine (MPW-260R). Afterwards, 5 ml of KI solution (5 %), 2 ml of glacial acetic acid and 2 ml of starch solution (2 %) were added to 5 ml of the prepared sample extract. Then, it was titrated with 0.001N KIO3 solution. Finally, the vitamin C content (mg/100 g) was estimated using the following formula-

$$\text{Vitamin C (mg/100 g)} = \frac{T \times F \times V \times 100}{v \times W}$$

Where, T = titrated volume of 0.001N $KIO_3$ (ml)
F = 0.088 mg of ascorbic acid per ml of 0.001 N $KIO_3$
V = total volume of sample extracted (ml)
v = volume of the extract (ml) titrated with 0.001 N $KIO_3$
W= sample weight (g).

### 2.6.5. Mineral compositions

The mineral content (Na, K, Ca and Mg) was determined by using device and method of an atomic absorption spectrophotometer (AAS), following procedures described by AOAC (1984) and Morshed et al. (2021) with some modifications. In this aspects, 0.5 g of sample powder was mixed with 5ml of the mixture of HNO and $HClO_4$ (5:1) and digested through a sand bath for 3-4 h. Digested extract was filtered and added distilled water up to the final volume of 100 ml. Then, 10 ml sample extract was taken and diluted to 50 ml with distilled water. Afterwards, the absorbance was measured through AAS (atomic absorption spectrophotometer; model-PinAAcle 900H; PerkinElmer) and the mineral concentration was measured according to the following formula:

$$\% \text{ Mineral} = \frac{\text{Sample reading x Final volume x Dilution factor}}{\text{Sample weight}}$$

### 2.7. Bioactive compounds

Determination of bioactive compounds was carried out in accordance with the standard methods using methanolic extracts of flesh and rind sample (Mohammed et al., 2020). For preparing methanolic extract, 1g of sample was immersed in 25 ml of methanol in a test tube and placed in a water bath at 30°C for two and half hour. Then the sample was centrifuged at 6000 rpm for 15 minutes and the supernatant was filtered and stored at 4°C in a refrigerator for further use.

### 2.7.1. Total phenolic content

Total phenolic content (TPC) was quantified spectrophotometrically using the Folin-Ciocalteu procedure (Mohammed et al., 2020) with some modifications. During analyses, 0.5 ml of methanolic extract was mixed with 2.5 ml of the Folin-Ciocalteu reagent and 2ml of 7.5% sodium carbonate. The resultant mixture was then incubated at 30°C for 1 hour in dark condition followed by measuring absorbance at 760 nm using spectrophotometer (PD-303 UV Spectrophotometer; APEL Co.) against methanol blank. Different concentration of gallic acid was used to calculate the standard curve and the result was expressed as mg gallic acid equivalent (GAE) per 100 g of fresh weight.

### 2.7.2. Total flavonoid content

The spectrophotometer assay for the quantitative determination of total flavonoid content (TFC) was carried out by the aluminum chloride colorimertic method (Pourmorad et al., 2006) with some modifications. Here, 100µl of methanolic extract at an appropriate dilution was mixed with 100 µl of 10% (w/v) $AlCl_3$ and 100 µl of 1M potassium acetate. Then the mixture was incubated at room temperature in dark condition for 40 minutes followed by the measurement of absorbance at 420 nm using spectrophotometer against the methanol blank. Total flavonoid was quantified from the quercetin standard calibration curve and expressed as mg quercetin equivalent (QE) per 100 g fresh sample.

*2.8. Statistical Analysis*

All analyses were performed in triplicate following randomized complete block design (RCBD) and the average mean data were evaluated by ANOVA (analysis of variance) test and the means were compared following Duncan multiple range (DMRT) test to determine the significant differences ($P < 0.05$ using R software (4.0 version). In addition, correlation matrix, cluster analysis, principal component analysis (PCA) and biplot analysis were also performed by using GGally, agricolae, Factoextra, Corrplot packages of R program.

**3. Results and discussion**

*3.1. Fruit physical attributes*

The quality of fruit is a major aspect from the point of view of consumers which is determined by both physical and biochemical properties. Physical properties include size, shape and color of fruit which are immediate preference in the market for visual attractions (Usharani et al., 2022). Table 1 revealed that the studied mini watermelon genotypes expressed noticeable variations for their fruit morphological characteristics. In terms of fruit shape, it was found different among the mini watermelon genotypes. Three different types of fruit shapes were noticed during evaluation viz. round ($W_1$), oval ($W_2$ and $W_4$) and oblong ($W_3$ and $W_5$). Watermelon fruit shape can be elongated, oval, round or oblong based on the fruit length to width ratio (Lou and Wehmer, 2016).

Divergent rind color and patterns in watermelon are preference of consumers that makes it commercially important and considerable importance has also been given to its esthetic value (Kayesh et al., 2013).The color of the watermelon rind generally varies from light to dark green *(*solid or striped) and yellow (Gusmini and Wehner, 2007; Dou et al., 2018). In our study, fruits of $W_1$, $W_2$ and $W_3$ genotype had blackish, light and dark green rind, respectively while rest of the two genotypes ($W_4$ and $W_5$) showed combination of deep and light green color (light green with dark green stripe).

Watermelon flesh color is another vital appearance quality closely related to consumers' preferences (Yuan et al., 2021). Watermelon flesh colors include coral red, scarlet red, canary yellow, orange, and white flesh. These different colors of watermelon not only provide visual diversity but are important from a nutritional perspective, as they are based on the carotenoid composition and content (Song et al., 2023). Among the tested genotypes, fruits of $W_1$, $W_4$ and $W_5$ genotypes had red flesh; yellow color was noticed in $W_2$ whereas $W_3$ had yellowish orange flesh. Different type and content of carotenoids contributed to this variation in flesh color of watermelon (Song et al., 2023).

Fruit flesh texture properties, especially flesh firmness, influence sensory quality and affect taste, flavor, and the shelf life of ripened watermelon fruit (Gao et al., 2020; Sun et al., 2020). A hard flesh results in reduced juice and poor flavor, whereas a soft flesh shows poor eating quality, storage and reduced shelf life (Gao et al., 2020). During evaluation, juicy compact flesh texture was recorded in $W_2$, $W_3$ and $W_5$ genotype. Meanwhile, crispy and sandy compact fleshes were noticed for $W_1$ and $W_4$ genotype, respectively.

**Table 1**
Fruit morphological attributes of five mini watermelon genotypes.

| Genotype | Fruit shape | Rind color | Flesh color | Flesh texture |
|---|---|---|---|---|
| $W_1$ | Round | Blackish green | Red | Crispy |
| $W_2$ | Oval | Light green | Yellow | Juicy compact |
| $W_3$ | Oblong | Dark green | Yellowish orange | Juicy compact |
| $W_4$ | Oval | Light green with dark green stripe | Red | Sandy compact |
| $W_5$ | Oblong | Light green with dark green stripe | Red | Juicy compact |

Physical traits of mini watermelon fruits in terms of size, rind thickness and weight showed significant variation ($p < 0.05$) among them (Table 2).

Fruit length along with diameter is a good indicator of better quality watermelon. The measure of fruit equatorial diameter decides the shape of the fruit. Lesser its value, more is the fruit shape towards the cylindrical shape (Tamil selvi et al., 2012). In the current study, the length of the fruit was measured within the range of 15.06 to 22.09 cm and the longest fruit was produced by $W_5$ genotype whereas $W_1$ genotype had the shortest one (15.06 cm) exhibiting statistical harmony with $W_2$ followed by $W_4$ genotype. In terms of fruit diameter, it was also found maximum (12.93 cm) in $W_5$ genotype followed by $W_1$ while fruit of $W_2$ possessed the minimum diameter (9.59 cm) that was statistically at par with $W_3$ genotype. In a study, Sari et al. (2016) found fruit length varied from 14.53 to 23.67 cm and fruit diameter from 11.71 to 17.94 cm in 38 mini watermelon lines. These findings were in close conformity with our results.

Rind thickness in watermelon fruits is an important feature for packaging. Fruits with very thin rind require a greater care in transport to the final destination and have a shorter shelf life, an undesirable characteristic for both the trader and the final consumer (Rouphael et al., 2010). Present study revealed that the thickest rind (1.57 cm) was found in $W_1$ genotype and the thinnest (0.74 cm) was traced in the fruit of $W_4$ genotype. Our range of rind thickness was found higher than that of reported by Sari et al., (2016).

Fruit weight in watermelon production is an important descriptor of fruit type, although it can also be considered as a yield component (Gusmini and Wehner, 2007). Average fruit weight of the studied genotypes was recorded within 2.16 to 3.79 kg and the lightest fruit (2.16 kg) was harvested from $W_3$ genotype while rest of the genotypes produced fruits with statistically similar weight of which $W_4$ had heavier fruit (3.79 kg). This finding was in consistent with earlier work by Sari et al., (2016) who reported that the fruit weight was remarkably varied within 1.21 to 3.59 kg among the mini watermelon lines.

**Table 2**
Fruit length, diameter, rind thickness and fruit weight of mini watermelon genotypes.

| Genotype | Fruit length (cm) | Fruit diameter (cm) | Rind thickness (cm) | Average fruit weight (kg) |
|---|---|---|---|---|
| $W_1$ | $15.06 \pm 0.59^{cx}$ | $11.77 \pm 0.89^{ab}$ | $1.57 \pm 0.04^a$ | $3.19 \pm 0.48^a$ |
| $W_2$ | $15.11 \pm 1.48^c$ | $9.59 \pm 0.55^c$ | $0.96 \pm 0.04^c$ | $3.15 \pm 0.67^a$ |
| $W_3$ | $18.37 \pm 0.22^b$ | $10.19 \pm 0.20^c$ | $1.27 \pm 0.11^b$ | $2.16 \pm 0.17^b$ |
| $W_4$ | $17.33 \pm 2.54^{bc}$ | $11.52 \pm 0.11^b$ | $0.74 \pm 0.92^d$ | $3.79 \pm 0.19^a$ |
| $W_5$ | $22.09 \pm 0.36^a$ | $12.93 \pm 0.81^a$ | $0.88 \pm 0.04^c$ | $3.26 \pm 0.57^a$ |

$^x$Data presented as means ± standard deviation in each column followed by the same letter(s) are not significantly different at p<0.05 as determined by Duncan Multiple Range test (DMRT) using the R software.

*3.2. Nutritional attributes*

Fruits of different mini watermelon genotypes detected significant differences (p < 0.01) among them in terms of nutritional abundance.

*3.2.1. Total soluble solids (TSS)*
Flesh sweetness is one of the prime internal as well as eating quality determining factors of fresh watermelon fruit which is related to the total soluble solids (TSS) (Yativ et al., 2010; Liu et al., 2013; Yau et al, 2010). By international standards, the fruit can be classified according to the refractometric index when measured at the midpoint of the fruit in the equatorial section. Any watermelon with ≥ 8 °Brix at the center of the flesh is sufficiently ripe and considered good internal quality and that with 10 °Brix is of very good internal quality (Kyriacou et al., 2018). Among the genotypes, $W_1$ was found to be the sweetest genotype with the highest TSS content (10.79 °Brix) whereas $W_4$ genotype had the lowest value (9.08 °Brix) showing statistical unity with $W_5$ (Figure 1). Hence, $W_1$, $W_2$ and $W_3$ genotype with TSS above 10 °Brix in the current study could be considered as fruits with very good internal quality. Sari et al. (2016) reported TSS within the range of 6.74 to 11.45 °Brix in their mini watermelon lines and our result showed values within that range. However, Schultheis et al. (2007) found TSS ranging from 10.6 to 12.0°Brix in 26 seedless mini watermelon varieties which was higher than our estimation.

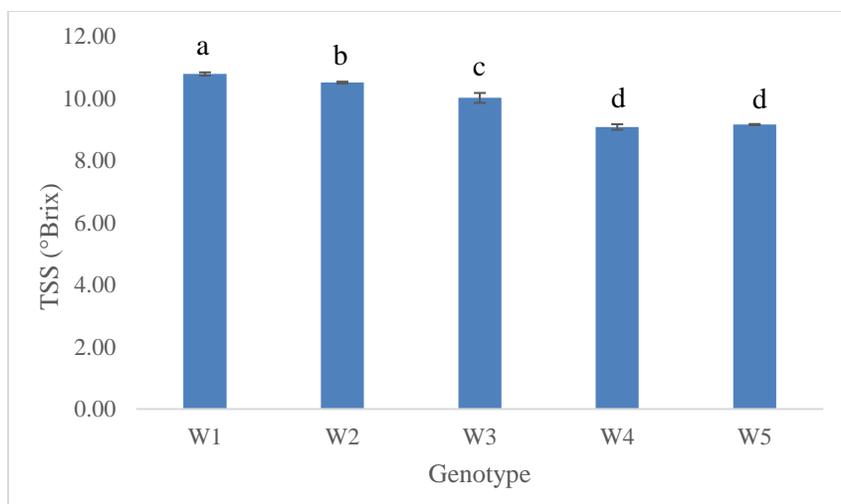

**Fig. 1:** TSS of different mini watermelon fruits. [Bars under each parameter with the same letter (s) are not significantly different at P < 0.01; error bars indicate standard deviation].

### 3.2.2. Total and reducing sugar content

Sugar contents in flash and rind samples of watermelon fruit varied significantly ($p < 0.01$) among the five genotypes (Table 3). $W_2$ genotype performed well with the maximum value of total sugar (18.95 mg/100 g) and reducing sugar content (17.33 mg/100g) in its flesh. Contrarily, $W_3$ genotype contained 8.95 and 5.01 mg/100 g of total and reducing sugar content, respectively which was found to be the minimum in case of flesh. Meanwhile, rind of $W_5$ genotype possessed maximum amount of total sugar (17.33 mg/100g) and reducing sugar content (9.16 mg/100g) whereas $W_4$ had the minimum content of sugar (3.80 and 1.99 mg/100 g total and reducing, respectively). Our values for reducing sugar of flesh were lower than that of Soumya and Rao (2014).

**Table 3**
Total and reducing sugar content of various mini watermelon genotypes.

| Genotype | Total sugar (mg/100 g) | | Reducing sugar (mg/100 g) | |
| --- | --- | --- | --- | --- |
| | Flesh | Rind | Flesh | Rind |
| $W_1$ | $12.32 \pm 0.08^{c\ x}$ | $10.61 \pm 0.39^b$ | $7.18 \pm 0.20^c$ | $3.62 \pm 0.39^d$ |
| $W_2$ | $18.95 \pm 0.45^a$ | $11.06 \pm 0.30^b$ | $9.24 \pm 0.42^a$ | $7.52 \pm 0.16^b$ |
| $W_3$ | $8.95 \pm 0.61^d$ | $8.97 \pm 0.64^c$ | $5.01 \pm 0.11^e$ | $4.42 \pm 0.14^c$ |
| $W_4$ | $12.10 \pm 1.01^c$ | $3.80 \pm 0.31^d$ | $5.63 \pm 0.38^d$ | $1.99 \pm 0.06^e$ |
| $W_5$ | $15.35 \pm 0.52^b$ | $17.33 \pm 0.58^a$ | $8.42 \pm 0.10^b$ | $9.16 \pm 0.17^a$ |

$^x$Data presented as means ± standard deviation in each column followed by the same letter(s) are not significantly different at p<0.01 as determined by Duncan Multiple Range test (DMRT) using the R software.

### 3.2.3. β carotene

Carotenoids such as β-carotene are an important dietary source of *vitamin A* (Haskell, 2012; Tang, 2012). Generally, higher beta carotene content would increase the nutritive value of the fruits (Venkatesan et al., 2016). Analysis of variance revealed significant difference ($p < 0.01$) among the five genotypes regarding β carotene content in the fruit flesh and rind (Table 4). β carotene content in the flesh and rind of the tested mini watermelons fluctuated within the range of 0.03 to 0.17 and 0.01 to 0.23 mg/100 g, respectively and $W_3$ genotype was found enriched with the highest amount in both flesh (0.17 mg/100 g) and rind (0.23 mg/100 g). In contrast, flesh of $W_5$ genotype had the least amount (0.03 mg/100 g) of β carotene whereas rind of $W_4$ contained the lowest value (0.01 mg/100 g) having statistical affinity with that of $W_1$ genotype. Quite similar values in watermelon flesh ranged from 0.1 to 2.1 mg per kg fresh weight were obtained by Tlili et al. (2011). Furthermore, our obtained results were lower than those of Tlili et al. (2023) who recorded the β-carotene level in watermelon cultivars ranging from 1.54 to 10.39 mg per kg fresh weight.

### 3.2.4. Vitamin C

Ascorbic acid is an active form of vitamin C that can impart a sour taste and its amount varies in different species of fruits and vegetables (Manchali et al., 2021; Soumya and Rao, 2014). It is of much importance from nutrition point of view due to its antioxidant property (Dhillon et al., 2019). The estimated value of vitamin C was found to be the highest in the flesh (32.85 mg/100 g) and rind (29.70 mg/100 g) of $W_5$ and $W_1$ genotype, respectively (Table 4). Oppositely, the lowest amount was registered in the fruit of $W_2$ genotype (10.40 mg/100 g in flesh and 10.46 mg/100 g in rind). These findings were concurring with that reported by Tlili, et al. (2023) who obtained total vitamin C of watermelon flesh within the range 113.43 to 241.16 mg kg−1 fresh weight. The observed variability might be ascribed to genotypic differences, applied agricultural practices, degree of maturation at harvest and post-harvest handling (Leskovar et al., 2004; Tlili et al., 2011).

**Table 4**
β carotene and C content in flesh and rind of mini watermelon fruits.

| Genotype | β carotene (mg/100 g) | | Vitamin C (mg/100 g) | |
|---|---|---|---|---|
| | Flesh | Rind | Flesh | Rind |
| $W_1$ | $0.15 \pm 0.00^{bx}$ | $0.02 \pm 0.00^d$ | $15.19 \pm 0.04^d$ | $29.70 \pm 0.57^a$ |
| $W_2$ | $0.06 \pm 0.00^c$ | $0.07 \pm 0.00^c$ | $10.40 \pm 0.03^e$ | $10.46 \pm 0.09^d$ |
| $W_3$ | $0.17 \pm 0.00^a$ | $0.23 \pm 0.01^a$ | $18.46 \pm 0.02^b$ | $16.03 \pm 0.03^c$ |
| $W_4$ | $0.14 \pm 0.00^b$ | $0.01 \pm 0.01^d$ | $15.63 \pm 0.06^c$ | $16.14 \pm 0.22^c$ |
| $W_5$ | $0.03 \pm 0.00^d$ | $0.14 \pm 0.01^b$ | $32.85 \pm 0.25^a$ | $20.24 \pm 0.52^s$ |

$^x$Data presented as means ± standard deviation in each column followed by the same letter(s) are not significantly different at $p<0.01$ as determined by Duncan Multiple Range test (DMRT) using the R software.

### 3.2.5. Mineral compositions

Dietary mineral elements are crucial for good and balanced human nutrition. They support a wide variety of bodily functions such as; building and maintaining healthy bones and teeth, keeping the muscles in shape and improving the functions of the heart and brain (Jéquier and Constant, 2010). Table 5 shows that significant variability ($p < 0.01$) the five mini watermelon genotypes were observed for their mineral contents. Na content varied within the range of 0.02 to 0.10% for flesh and 0.04 to 0.10% for rind of mini watermelon fruits. In case of flesh, significantly the highest content of Na (0.10%) was noted in $W_4$ genotype and the lowest value was recorded in $W_3$ (0.02%). Besides, rind of both $W_3$ and $W_6$ genotype had the highest amount of Na of 0.10%. Conversely, the least content of Na was recorded in the rind of $W_2$ genotype which was statistically alike with $W_1$ genotype.

K content of the flesh and rind was recorded maximum (1.39%) in $W_4$ and $W_3$ genotype, respectively whereas the minimum level was noticed in both flesh (0.67%) and rind (1.03%) of $W_1$ genotype.

Again, among the genotypes, percentage of Ca in the flesh was measured maximum (0.28%) in $W_2$ genotype having statistical consistency with that of $W_4$ genotype while minimum amount (0.16%) was determined in $W_1$ and $W_3$ genotype. On the other hand, fruit rind with the highest content of Ca (0.28%) was observed in $W_5$ genotype and the least value (0.15%) was detected in $W_4$ genotype which had statistical unity with $W_1$ genotype.

The result also showed that fruit with maximum content of Mg in its flesh was harvested from $W_5$ followed by $W_3$ genotype whereas minimum level (0.28%) was observed in $W_1$ expressing statistical parity with $W_2$ genotype. However, fruit rind with the highest (0.38%) and lowest (0.2%) amount of Mg belonged to $W_3$ and $W_1$ genotype, respectively. Feizy et al. (2020) also measured 468.00 ± 0.12, 164.48 ± 0.20, 2074.00 ± 10.00, 53.59 ± 0.10 mg/100 g calcium, magnesium, potassium and sodium, respectively in watermelon rind which were nearly similar to the values of present study.

**Table 5**
Mineral contents of flesh and rind of different mini watermelon genotypes fruits.

| Genotype | Na (%) | | K (%) | | Ca (%) | | Mg (%) | |
|---|---|---|---|---|---|---|---|---|
| | Flesh | Rind | Flesh | Rind | Flesh | Rind | Flesh | Rind |
| $W_1$ | 0.05 ± 0.00$^{dx}$ | 0.05 ± 0.00$^c$ | 0.67 ± 0.00$^e$ | 1.03 ± 0.01$^d$ | 0.16 ± 0.00$^c$ | 0.16 ± 0.01$^d$ | 0.28 ± 0.04$^c$ | 0.20 ± 0.00$^e$ |
| $W_2$ | 0.07 ± 0.00$^c$ | 0.04 ± 0.00$^c$ | 1.30 ± 0.00$^c$ | 1.13 ± 0.01$^c$ | 0.28 ± 0.00$^a$ | 0.20 ± 0.01$^c$ | 0.31 ± 0.00$^c$ | 0.34 ± 0.01$^b$ |
| $W_3$ | 0.02 ± 0.00$^e$ | 0.10 ± 0.00$^a$ | 1.09 ± 0.00$^d$ | 1.37 ± 0.01$^a$ | 0.16 ± 0.01$^c$ | 0.23 ± 0.01$^b$ | 0.36 ± 0.01$^{ab}$ | 0.38 ± 0.01$^a$ |
| $W_4$ | 0.10 ± 0.00$^a$ | 0.06 ± 0.01$^b$ | 1.39 ± 0.00$^a$ | 1.16 ± 0.01$^b$ | 0.27 ± 0.00$^a$ | 0.15 ± 0.01$^d$ | 0.35 ± 0.01$^b$ | 0.31 ± 0.01$^c$ |
| $W_5$ | 0.08 ± 0.00$^b$ | 0.10 ± 0.01$^a$ | 1.33 ± 0.00$^b$ | 1.15 ± 0.01$^{bc}$ | 0.23 ± 0.01$^b$ | 0.28 ± 0.01$^a$ | 0.39 ± 0.00$^a$ | 0.25 ± 0.01$^d$ |

$^x$Data presented as means ± standard deviation in each column followed by the same letter(s) are not significantly different at $p<0.01$ as determined by Duncan Multiple Range test (DMRT) using the R software.

*3.3. Bioactive compounds*

Fruits produce a wide array of secondary metabolites, which perform essential physiological and biochemical functions. These metabolites are also of utmost importance in fruit quality from the point of view of consumer acceptability, affecting the color/appearance and the flavor, and their implication in fruit nutritional characteristics (Sanchez-Ballesta et al., 2022). Flavonoids and phenolic acids are the most important groups of secondary metabolites and bioactive compounds in plants (Kim et al., 2003).

*3.3.1. Total phenolic content*

Phenolic compounds have gained much attention due to their antioxidant activities and free radical scavenging abilities, and they have potential beneficial implications for human health (Soumya and Rao (2014). With respect to the total phenolic content (TPC), significant differences ($p < 0.01$) among the watermelon genotypes were evident (Table 6). Total phenol content was measured the utmost in the flesh of W3 (107.08 mg GAE/100 g) and rind of W1 genotype (89.74 mg GAE/100 g). Meanwhile, the lowest content belonged to the flesh of W4 (8.44 mg GAE /100 g) and rind of W genotype (5.7644 mg GAE/100 g). This result was comparable to the findings of Tlili, et al. (2023) who reported the total phenolic content in flesh of watermelon cultivars varied from 79.55 to 243.51 mg GAE per kg fresh weight. Additionally, the amount of total phenolic compounds in peel measured by Feizy et al. (2020) was found 2473.45 mg gallic acid equivalent (GAE)/100 g. The varied trend of phenolic content was probably due to the different degree of the biosynthetic pathways of these compounds affected during ripening and also might be due to genetic and environmental factors (Kolayli et al., 2010).

*3.3.2. Total flavonoid content*

Flavonoids are the phenolic compounds which possess free radical scavenging activity and are linked to multiple health benefits including antioxidant, anti-carcinogenic and anti-inflammatory (Rocha et al., 2005). It is well-known that flavonoid contribute to nutritional value and food quality in terms of modifying color, taste, aroma, and flavor (Panche et al., 2016). As depicted in Table 6, total flavonoid content (TFC) accumulated maximum in the flesh of $W_3$ (18.37 mg QE/100 g) and rind of $W_6$ genotype (17.17 mg QE/100 g). On the reverse side, minimum amount was registered in the $W_1$ genotype (14.87 mg QE/100 g in flesh and 14.01 mg QE/100 g in rind). These differences in total flavonoid content might be due to the different genotypes of watermelons that were analyzed.

**Table 6**

TPC and TFC of flesh and rind of mini watermelon genotypes fruits.

| Genotype | TPC (mg GAE/100 g) | | TFC (mg QE/100 g) | |
| --- | --- | --- | --- | --- |
| | Flesh | Rind | Flesh | Rind |
| $W_1$ | $15.81 \pm 0.08^{dx}$ | $89.74 \pm 0.51^a$ | $14.87 \pm 0.14^d$ | $14.01 \pm 0.11^c$ |
| $W_2$ | $20.96 \pm 0.20^b$ | $13.82 \pm 0.27^d$ | $17.26 \pm 0.04^b$ | $16.55 \pm 0.60^b$ |
| $W_3$ | $107.08 \pm 0.58^a$ | $55.48 \pm 0.55^c$ | $18.37 \pm 0.03^a$ | $16.69 \pm 0.15^b$ |
| $W_4$ | $8.44 \pm 0.13^e$ | $61.27 \pm 0.28^b$ | $17.06 \pm 0.06^c$ | $16.62 \pm 0.64^b$ |
| W₅ | 18.87 ± 0.16$^c$ | 5.76 ± 0.47$^e$ | 16.99 ± 0.06$^c$ | 17.17 ±0.18$^a$ |

$^x$Data presented as means ± standard deviation in each column followed by the same letter(s) are not significantly different at p<0.01 as determined by Duncan Multiple Range test (DMRT) using the R software.

*3.4. Multivariate analysis*

Pearson's correlation matrix expressed the interrelationship among the studied 25 variables related to the mini watermelon fruit quality (Figure 2A). From this analysis, it was revealed that fruit size (length and diameter) had very weak correlation with fruit weight which explains that fruit weight of melon didn't increase with the increase of fruit length and breadth. While considering the vitamin content, fruit size exhibited moderate to strong correlation with vitamin C but almost no correlation with β carotene describing that size enhancement promoted the vitamin C content in watermelon, not β carotene content. Among the biochemical and bioactive compounds, TSS and TPC had strong negative correlation with fruit size and weight, respectively which denotes a reverse relation between fruit size and these TSS and TPC of watermelon. Whereas, TFC had very weak positive and sugar content showed almost no correlation with fruit morphology. Mineral contents in the fruit flesh expressed weak to moderate correlations with fruit size except for Na and Mg which had strong positive relation with fruit weight and fruit length, respectively. Regarding the rind characters, fruit size (length and diameter) had moderate to strong positive correlation with all the physio-chemical, functional and mineral properties of rind except for TPC, vitamin C and Mg content. Whereas, rind thickness as well as fruit weight expressed very weak or null correlations with the studied rind properties. Such diversified relationship among the fruit physical and biochemical properties displayed wide variability among the genotypes.

Heatmap with dendrogram cluster prepared using the 25 studied dependent variables depicted 2 main clusters (Figure 2B). Cluster 1 included the variables like fruit diameter, vitamin C (rind), β carotene (flesh), TPC (rind), rind thickness and TSS which were closely related to each other. Meanwhile, cluster 2 contained rest of the variables which were further grouped into two sub clusters. Average fruit weight, fruit length, Na (flesh), Ca (flesh and rind), reducing sugar and total sugar content (flesh and rind) formed sub cluster 1 while TFC (flesh and rind), Mg (flesh and rind), K (flesh and rind), TPC (flesh), β carotene (rind), vitamin C (flesh) and Na (rind) were included in sub cluster 2.

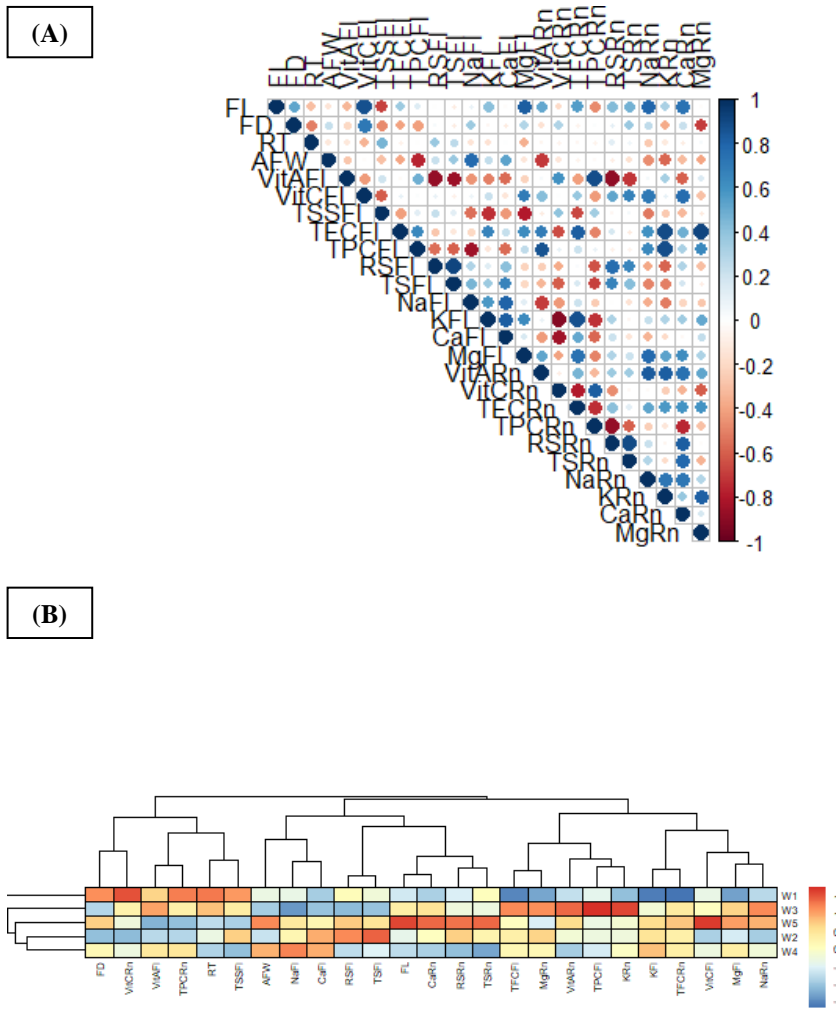

**Fig. 2.** (A) Correlation coefficient for variables related to fruit morphological and nutritional quality in mini watermelon; (B) Distribution of 25 variables into two major clusters with Heatmap.

Principal component analysis (PCA) simplifies the wide range of data by transforming the number of correlated variables into smaller number of variables. As observed, the first two principal components (dimension 1 and dimension 2) contributed enough to explain maximum (64.8%) of the pattern variations. Among the variables, TPC (flesh and rind), β carotene (rind) and K content (rind) were noticed as strong; total sugar content (rind) and fruit diameter were found as less contributing and rest of the parameters were noticed as intermediate contributing variables (Figure 3A).

As seen in Figure 3B, rind thickness, average fruit weight, TSS, β carotene of flesh, vitamin C and TPC of rind were positively correlated with this variation considering the PC1while PC2 were found positively correlated with fruit length, rind thickness, TSS, vitamins, TPC, TFC, flesh Mg and rind minerals. Among these variables, rind thickness, TSS, β carotene (flesh), vitamin C (rind) and TPC (rind) were commonly found as positive loading factors in both dimensions. These variables were, therefore, the most contributing factors indicating differences among the genotypes and the importance of selecting the proper genotypes to provide high-quality values.

When we consider the PCA-biplot, we have observed that the five mini watermelon genotypes were found in five distinct positions (Figure 3C). $W_1$ and $W_3$ were found to locate at the positive side of dimension 1 and dimension 2, respectively. Among the rest three genotypes, $W_2$ and $W_4$ positioned very near to each other showing close statistical similarity. This PCA findings were further clarified by cluster dendrogram analysis which showed that five watermelon genotypes were grouped into two main clusters where Cluster 1 contained only $W_1$ genoytype showing distinctness to the other genotypes and Cluster 2 was further divided into two sub clusters showing W3 Genotype in one subcluster and rest three genotypes remained in another sub cluster (Figure 3D).

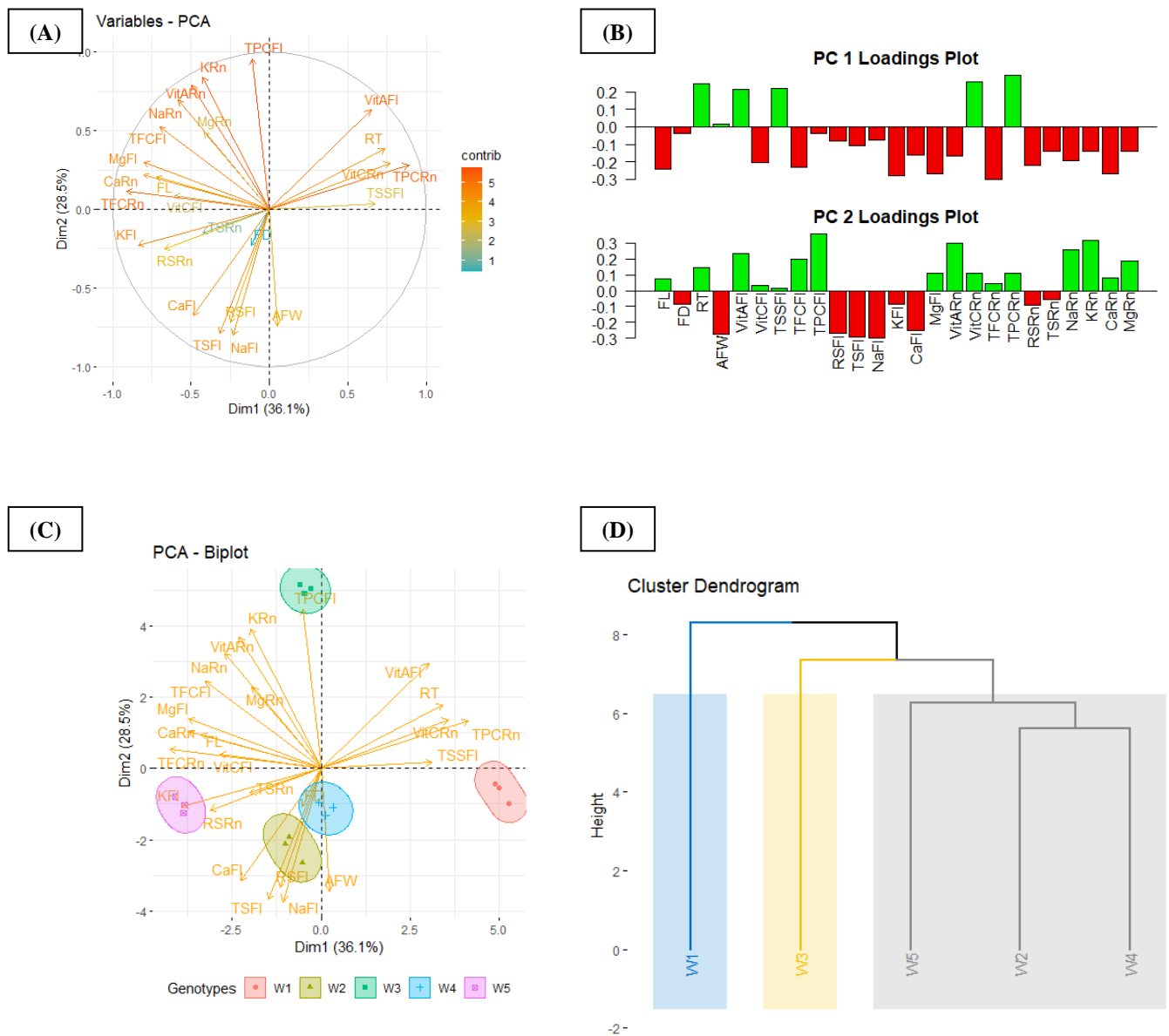

**Fig. 3.** (A) PCA of the variables showing their major contribution; (B) factor loadings for the first two principal components (Dim 1 and Dim 2); (C) PCA–biplot analysis representing the performance of genotypes regarding quality parameters; (D) Dendrogram categorizing the accessions according to their similarities.

## 4. Conclusion

From the results of this comparative study, BARI watermelon 1 ($W_1$) and L-32468 ($W_3$) genotypes showed to be promising in terms of fruit quality. Therefore, they could be grown to meet the market demands for mini watermelons with good quality fruit and would help breeders and other researchers to use them for mini watermelon improvement. Besides, being potential source of bioactive compounds, rind of $W_1$ and $W_5$ genotype could be considered as promising functional ingredients for food and industrial usage.

**Conflicts of interest**

The authors declare that they have no conflicts of interest.